\documentclass[mathleft
]{an}
\usepackage{graphicx}
\usepackage{times}
\begin{document}

\Pagespan{789}{}
\Yearpublication{2012}%
\Yearsubmission{2012}%
\Month{11}%
\Volume{999}%
\Issue{88}%

\title{Attempts to reproduce the rotation profile of the red giant KIC~7341231 observed by {\it Kepler}}

\author{T. Ceillier\inst{1} 
\and P. Eggenberger\inst{2}
\and R.~A. Garc\'\i a\inst{1}
\and S. Mathis\inst{1}
}
\titlerunning{Attempts to reproduce the rotation profile of the red giant KIC~7341231}
\authorrunning{Ceillier, Eggenberger, Garc\'\i a \& Mathis}
\institute{
Laboratoire AIM, CEA/DSM Ð CNRS - Univ. Paris Diderot Ð IRFU/SAp, Centre de Saclay, F-91191 Gif-sur-Yvette Cedex, France
\and 
Observatoire de Gen\`eve, Universit\'e de Gen\`eve, 51 chemin des Maillettes, CH-1290, Sauverny, Suisse
}

\received{15 August  2012}
\accepted{September 2012}
\publonline{later}

\keywords{stars: evolution -- stars: oscillations -- stars: rotation}

\abstract{
{Thanks to the asteroseimic study of the red giant star KIC~7341231 observed by {\it Kepler}, it has been possible to infer its radial differential rotation profile (Deheuvels et al. 2012). 
This opens new ways to constrain the physical mechanisms responsible of the angular momentum transport in stellar interiors by directly comparing this radial rotation profile 
with the ones computed using stellar evolution codes including dynamical processes. In this preliminary work, we computed different models of KIC~7341231 with the Geneva stellar evolution code that includes {transport mechanisms due to a} shellular rotation {and the associated large-scale meridional circulation and shear-induced turbulence}. Once the global parameters of the star had been established, we modified some of the model's input parameters in order to understand their effects on the predicted rotation profile of the modeled star. As a result, we find a discrepancy between the rotation profile deduced from asteroseismic measurements and the profiles predicted from models including shellular rotation {and related meridional flows and turbulence. This indicates that a most powerful mechanism is in action to extract angular momentum from the core of this star.}}}

\maketitle

\section{Introduction}
Thanks to space missions such as CoRoT (Baglin et al. 2006) and {\it Kepler} (Borucki et al. 2010; Koch et al. 2010), we have now access to long high-precision light curves for many stars exhibiting solar-like oscillations at different evolutionary stages, from the main sequence (e.g. Appourchaux et al. 2008; Garc\'\i a et al. 2009; Ballot et al. 2011 ;Chaplin et al. 2011) till more evolved ones such as subgiants (Deheuvels et al. 2010; Campante et al. 2011; Mathur et al. 2011, 2012b; Metcalfe et al. 2012) and red giants (e.g. de Ridder et al. 2009; Bedding et al. 2010; Mosser et al. 2010, 2011; Huber et al. 2011).
The study of these light curves through asteroseismology has enabled us to probe the general properties of the interior of thousands of red giants (e.g. Kallinger et al. 2010; Bedding et al. 2011;  Basu et al. 2011; Mosser et al. 2012a,b), while some of them have already been studied with a greater detail (e.g. Metcalfe et al. 2010; Beck et al. 2011; di Mauro et al. 2011; Mathur et al. 2012a).
The long time series recorded by the space instrumentation have allowed us to study the dynamics of stars. Indeed, the existence of surface magnetic activity allowed us to constraint the surface rotation (e.g. Barban et al. 2009; Mosser et al. 2009; Mathur et al. 2010; Fr\"ohlich et al. 2012) as well as to measure activity cycles (Garc\'\i a et al. 2010). Moreover, the analysis of rotational splittings of the oscillation modes (e.g.  Ballot et al. 2011) allows to derive strong constraints on the internal rotation profiles of these sub giants and red giants as it has already been done for the Sun. 

Rotational splittings of mixed modes have then been precisely determined for the red giant KIC~8366239 (Beck et al. 2012). A theoretical analysis of these results shows that models of red giants including shellular rotation\footnote{The shellular rotation hypothesis, i.e. an angular velocity ($\Omega$) which depends only on the radius ($r$), has been introduced by Zahn (1992) and is assumed to be enforced by a strong horizontal turbulent transport in stably stratified stellar radiation zones.}  and the related large-scale meridional circulation and shear-induced turbulence (see Zahn 1992; Maeder \& Zahn 1998; Mathis \& Zahn 2004; Decressin et al. 2009 for a description of their interactions) predict steep rotation profiles, which are incompatible with these measurements of rotational splittings (Eggenberger et al. 2012).
This study also shows that meridional circulation and shear turbulence alone produce an insufficient internal coupling and that the efficiency of the needed additional mechanism for the internal transport of angular momentum during the post-main sequence evolution can be strongly constrained 
thanks to asteroseismic measurements.

Very recently, it has been possible to infer for the first time the internal rotation profile of a red giant thanks to precise asteroseismic mearurements obtained for the 
star KIC~7341231 from the core up to its external layers (Deheuvels et al. 2012). This red giant exhibits a much lower mass (about 0.84\,$M_{\odot}$) than the red giant KIC~8366239 observed by Beck et al. (2012) (which has
a mass of about 1.5\,$M_{\odot}$). KIC~7341231 has an effective temperature between 5470 K (Casagrande et al. 2010) and 5483 K (Ammons et al. 2006).
Its metallicity [Fe/H] is very low and ranges from -2.18 (Laird et al. 1988) to -0.79 (Ammons et al. 2006). 
The combination of this low metalicity, the very high proper motion (39.18 mas/yr in RA and 255.25 in DEC according to van Leeuwen 2007), and its high radial velocity (-269.16 km$\cdot$ s$^{-1}$, Latham et al. 2002) 
indicates that KIC~7341231 is a halo star. 
It is then particularly interesting to perform a direct comparison between the rotation profile inferred for KIC~7341231 and profiles predicted by theoretical
models to investigate whether the conclusions obtained by Eggenberger et al. (2012) about the efficiency of the internal transport of angular momentum in red giants are still valid 
for a low-mass and low-metallicity red giant like KIC~7341231. In this work we thus study, from a theoretical point of view, the rotation profile of the star KIC~7341231, which is located near the base
of the red giant branch.

The {\it Kepler} short-cadence photometric light curve was corrected following the procedures described by Garc\'\i a et al. (2011) of a year long and it was analyzed by Deheuvels et al. (2012). They obtained global seismic properties with a large separation of $\Delta\nu=28.9 \pm 0.2$ $\mu$Hz and a period spacing (for dipole g modes) of $\Delta \Pi 1=107.1 \pm 2.3$ s.

Moreover, thanks to the measurement of 40 individual eigenmodes of both acoustic and mixed natures,  it was possible to study the individual rotational splittings in the non-radial ones, making possible to derive the radial differential-rotation profile of the star. Deheuvels et al. (2012) have shown that the core of KIC~7341231 is rotating at a frequency $\Omega_c= 710 \pm 51$ nHz (averaged on the innermost 1.4\% of the stellar radius, corresponding to 17\% of the total mass) while its surface rotation is much slower with $\Omega_s < 150 \pm 19$ nHz.

\section{Modeling the star}
We used the Geneva stellar evolution code (Eggenberger et al. 2008) in which the effects of shellular rotation and related meridional flows and turbulence on stellar evolution (Eggenberger et al. 2010) have been implemented. In this work, as a first step, we do not take into account internal waves and magnetic fields.
All the models we computed start at the ZAMS.
We assumed a metallicity [Fe/H] = -1 dex and an initial helium abundance $Y_{\rm ini} = 0.260$.
We selected a stop point corresponding to when the model's large separation was equal to the observed one. We would then compare the model's period spacing with the observed one.
In order to best reproduce the observed surface rotation, we focused on models with a low initial rotational velocity on the ZAMS of v$_{\rm{ini}}= 2\,\rm{km \cdot s}^{-1}$.

\section{Results and discussion}
 In accordance with Deheuvels et al. (2012), we find that a model with a mass M = 0.84 M$_\odot$ reproduces both the observed large separation and period spacing at the same age T = 13.01 Gyr. Its evolutionary track is shown in Figure 1.
 
 \begin{figure}
   \includegraphics[width=80mm,height=105mm]{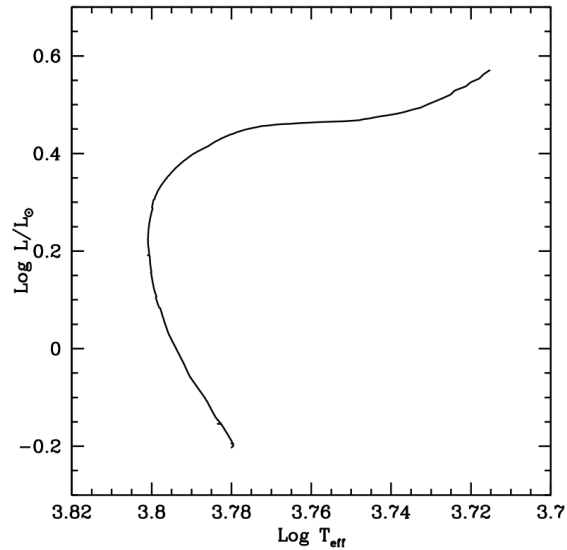}
   \caption{Evolutionary track of the selected model including rotation yielding to a mass of 0.84 M$_\odot$.}
   \label{label1}
 \end{figure}
 
We found that the obtained radial differential rotation profile was much steeper than the observed one, as can be seen in Figure 2. 
This is in good agreement with the results obtained by Eggenberger et al. (2012) for the more massive red giant KIC~8366239 
and this shows that shellular rotation and related meridional currents and turbulence alone also produces an insufficient internal coupling in the case of a low-mass and low-metallicity red giant like KIC~7341231 
to correctly reproduce its observed rotation profile. We recall that the same problem is found in the Sun for which the rotation profile obtained through helioseismology is almost flat until $0.2\,R_{\odot}$ (e.g. Garc\'\i a et al. 2007; Mathur et al. 2008), while modeling tends to give a steeper profile (e.g. Pinsonneault et al. 1989; Turck-Chi\`eze et al. 2010).

 \begin{figure}
   \includegraphics[width=80mm,height=100mm]{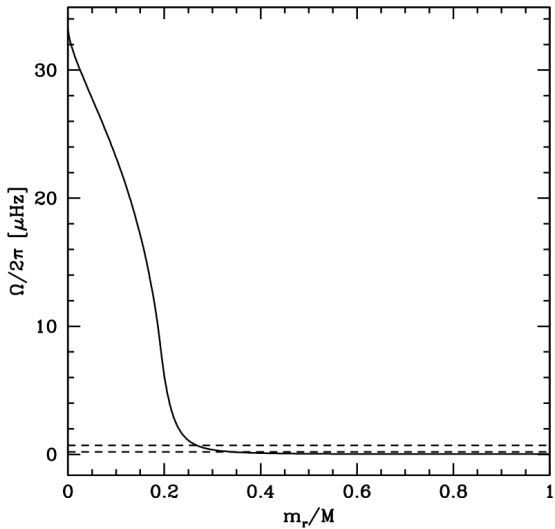}
   \caption{Rotation profile of the model at the end of the evolutionary track shown in Fig.~1 (solid line). The two dashed lines correspond to the core rotation rate and the surface rotation rate derived by Deheuvels et al. (2012).}
   \label{label1}
 \end{figure}

To investigate the effects of varying different modeling parameters on the rotation profile, we computed models with the same global properties (mass, v$_{\rm{ini}}$, metallicity...), but with/without magnetic braking during the main sequence, with/without atomic diffusion and also with different values of the mixing-length parameter and initial helium abundance. All these modifications implied slight changes in the obtained radial differential rotational profile, which are not sufficient to correctly reproduce the rotation profile deduced from asteroseismic measurements. 
This suggests that another and yet undetermined physical process is at work for the internal transport of angular momentum in addition to meridional circulation and shear turbulence as found
by Eggenberger et al. (2012) (see also Marques et al. 2012).
For the time being, it seems that the two best candidates for angular momentum transport from the core to the more external layers are internal gravity waves excited by the turbulent convective envelope (Zahn et al. 1997; Talon \& Charbonnel 2005 \& 2008; Mathis 2009; Mathis \& de Brye 2012) and fossil magnetic field and related MHD processes (Eggenberger et al. 2005; Mathis \& Zahn 2005; Zahn, Brun \& Mathis 2007; Strugarek et al. 2011). This two processes would tend to damp the differential rotation gradient and might explain the internal rotation profile of the Sun and of red giants.

\section{Conclusion}
By performing a direct comparison between the rotation profile inferred for the red giant KIC~7341231 and profiles predicted by stellar
models including shellular rotation, meridional circulation and shear turbulence, we found that theoretical rotation profiles are too steep to correctly reproduce the observed one.
This is in good agreement with the results obtained by Eggenberger et al. (2012) and this shows that this discrepancy is also found for
a low-mass and low-metallicity red giant star like KIC~7341231.

In addition to this preliminary study, it will be interesting to determine the efficiency of the unknown physical process
needed for the internal transport of angular momentum in order to correctly reproduce the observed rotation profile of KIC~7341231. This will allow a comparison
with the effective viscosity of $3 \times 10^{4}$\,cm$^2$\,s$^{-1}$ found for the more massive red giant KIC~8366239 (Eggenberger et al. 2012) and will give us some
clues about the variation of the efficiency of this unknown physical process with the stellar mass and metallicity. 

The present results illustrate the need to progress in our understanding and modeling of (magneto-) hydrodynamical processes at work in stellar interiors and 
to implement new physical processes into stellar evolution codes, such as internal gravity waves or magnetic fields. In such improvement of our vision of the dynamical evolution of stars, asteroseismology is then one of the most promising ways to get strong constraints as demonstrated here by the case of KIC~7341231.

\acknowledgements
TC, RAG and SM acknowledge the CNES for the support of the CoRoT and asteroseismic activities at the SAp, CEA/Saclay  and the CNRS/INSU PNPS support. TC and SM thank Geneva Observatory for its hospitality. PE was partly supported by the Swiss National Science Foundation.


\end{document}